\documentclass[10pt,journal]{IEEEtran} 
\usepackage{amsmath,amsfonts}
\usepackage{amssymb}
\usepackage{algorithmic}
\usepackage{graphicx}
\usepackage{booktabs}
\usepackage{comment}
\usepackage{caption}
\usepackage{subcaption}
\usepackage{nth}
\usepackage{textcomp}
\usepackage{xcolor}
\usepackage[disable]{todonotes}
\usepackage[colorlinks=true, urlcolor=blue, linkcolor=blue]{hyperref} 

\usepackage{tcolorbox} 
\tcbuselibrary{most} 

\usepackage[capitalise]{cleveref}
\crefformat{section}{#2\S#1#3} \crefformat{subsection}{#2\S#1#3}
\crefformat{subsubsection}{#2\S#1#3}
\crefformat{figure}{#2Fig. #1#3}
\crefformat{table}{#2Table #1#3}
\crefformat{algorithm}{#2Alg. #1#3}

\newboolean{showcomments}
\setboolean{showcomments}{false}
\ifthenelse{\boolean{showcomments}}
{ \newcommand{\mynote}[3]{
                \fbox{\bfseries\sffamily\scriptsize#1}
                {\small$\blacktriangleright$\textsf{\emph{\color{#3}{#2}}}$\blacktriangleleft$}}}
{\newcommand{\mynote}[3]{}}
\newcommand{\shrink}[1]{}
\definecolor{pink}{rgb}{1,0.2,0.7}
\definecolor{purple}{rgb}{0.7,0,0.9}
\newcommand{\yc}[1]{\mynote{Yun-Chih}{#1}{purple}}

\makeatletter
\patchcmd{\@maketitle}
  {\addvspace{0.5\baselineskip}\egroup}
  {\addvspace{-1.92\baselineskip}\egroup}
  {}
  {}
\makeatother

\begin{document}

\title{Search-in-Memory: Reliable, Versatile, and Efficient Data Matching in SSD's NAND Flash Memory Chip for Data Indexing Acceleration}

\author{Yun-Chih Chen,~\IEEEmembership{Member,~IEEE,}~Yuan-Hao~Chang,~\IEEEmembership{Fellow,~IEEE,\ }and ~Tei-Wei~Kuo,~\IEEEmembership{Fellow,~IEEE,}
\IEEEcompsocitemizethanks{
\IEEEcompsocthanksitem Y.-C. Chen is with the Department of Computer Science, TU Dortmund, Germany (E-mail: yunchih.chen@tu-dortmund.de).%
\IEEEcompsocthanksitem Y.-H. Chang is with the Institute of Information Science, Academia Sinica, Taipei 115, Taiwan (R.O.C.) (E-mail: johnson@iis.sinica.edu.tw).
\IEEEcompsocthanksitem T.-W. Kuo is with Delta Electronics, Department of Computer Science and Information Engineering, High Performance and Scientific Computing Center, and Center of Data Intelligence: Technologies, Applications, and Systems of National Taiwan University (E-mail: ktw@csie.ntu.edu.tw).%
}

\thanks{This paper has been accepted for presentation at the The International Conference on Hardware/Software Codesign and System Synthesis (CODES+ISSS) in September, 2024.  An extended abstract of this paper was presented in Design, Automation \& Test in Europe Conference \& Exhibition (DATE), 2024.}
}

\maketitle

\begin{abstract}
    To index the increasing volume of data, modern data indexes are typically stored on SSDs and cached in DRAM.  However, searching such an index has resulted in significant I/O traffic due to limited access locality and inefficient cache utilization. At the heart of index searching is the operation of filtering through vast data spans to isolate a small, relevant subset, which involves basic equality tests rather than the complex arithmetic provided by modern CPUs. We introduce the Search-in-Memory (SiM) chip, which demonstrates the feasibility of performing data filtering directly within a NAND flash memory chip, transmitting only relevant search results rather than complete pages. Instead of adding complex circuits, we propose repurposing existing circuitry for efficient and accurate bitwise parallel matching.  We demonstrate how different data structures can use our flexible SIMD command interface to offload index searches. This strategy not only frees up the CPU for more computationally demanding tasks, but it also optimizes DRAM usage for write buffering, significantly lowering energy consumption associated with I/O transmission between the CPU and DRAM. Extensive testing across a wide range of workloads reveals up to a 9X speedup in write-heavy workloads and up to 45\% energy savings due to reduced read and write I/O. Furthermore, we achieve significant reductions in median and tail read latencies of up to 89\% and 85\% respectively.
\end{abstract}

\newcommand{\opencmd}{\texttt{page-open}\ command}
\newcommand{\searchcmd}{\texttt{search}\ command}
\newcommand{\gathercmd}{\texttt{gather}\ command}
\newcommand{\cachecmd}{\texttt{page-open-cache}\ command}

\section{Introduction}
\subsubsection*{Challenges of indexing vast amount of data}
Data indexes, such as hash tables and trees, are fundamental for quickly retrieving relevant data from vast datasets. As the volume of data to be indexed explodes, the size of the index is growing significantly large. In many user-facing databases that  execute complex queries, index size can even surpass the data being indexed \cite{oltp16}.  Given that accessing an index invariably precedes any data retrieval, indexes are commonly pinned in-memory to boost performance. With the introduction of high-speed SSDs, even systems sensitive to latency-—those interfacing directly with users-—resort to storing indexes on SSDs and loading them into DRAM on-demand.  Upon loading an index block (for instance, a B-Tree's leaf node or a hash table's bucket) into DRAM, a subsequent scan through the memory page that contain arrays of candidate entry is necessary to find the matching one.  Such a parallel equality test is often accelerated with SIMD instructions.

As I/O can easily become the bottleneck, compression and data prefetching are common techniques employed to reduce I/O and hide latency. However, in many workloads indexes exhibit low compressibility, and decompression incurs overhead \cite{oltp16}. Moreover, prefetching can accelerate the replacement of loaded index blocks. In large-scale data systems, where the working set size far exceeds DRAM capacity and the accesses scatter widely, index blocks can be repetitively loaded and evicted from DRAM.  Even if all index blocks fit entirely in DRAM, they can still be evicted after context-switching to other processes that might also allocate memory.  Another pressing issue is the management of index updates. These updates not only require considerable buffering to mitigate the SSD's high write costs but also introduce multiple data versions that compete for the limited DRAM cache space with index reads, leading to increased I/O due to more frequent read cache misses.

To solve the I/O bottleneck, one can either increase DRAM capacity or I/O bandwidth.
However, both approaches bring substantial costs and power consumption. In environments where cost efficiency is as crucial as performance, the focus should not solely be on maximizing index retrieval's throughput but on enhancing the utility of the retrieved indexes.  Perhaps the best way is to fundamentally cut the amount of indexes that need to be transferred from the storage system.  

There have been numerous innovations in data structures aimed at optimizing data indexing and system-level optimizations, such as kernel bypassing, to maximize I/O bus utilization.  This paper takes a different approach, focusing on the core operation of data indexing: matching a query against a vast array of candidate entries.
Within the constraints of today's von Neumann architecture, this equality test operation occurs in the CPU only after transferring all candidate entries from storage. Yet, this operation, predominantly data-bound, does not require the complex arithmetic or control flows modern CPUs offer and could be executed by simpler hardware circuits.

This leads us to question whether equality tests could be integrated deeper into the storage system. While Processing-in-Memory (PiM) has been explored as a solution to the bottleneck between the CPU and DRAM, it does not address DRAM’s capacity scaling challenges. Conversely, a NAND-flash-memory-based solution offers higher energy efficiency and capacity.   In this paper, we explore this direction by introducing the Search-in-Memory (SiM) chip.

SiM is based on the architecture of existing TLC flash memory chip.  Instead of introducing a full-fledged hardware-based indexing solution, we aim to minimize hardware changes and use software to decompose complex indexing operations into simple hardware instructions, similar to the design philosophy of RISC CPUs.
We demonstrate how to minimally modify an existing flash memory chip to conduct equality tests directly in itself and send only the relevant results in response to a search request rather than the entire page to fundamentally reduce the I/O traffic.

In our experiment, we also demonstrate the performance characteristics of index search under various workloads, query distribution, and system constraints, as well as how SiM can improve system efficiency by reducing I/O transmission and increasing cache utilization.
 We make the following contributions:

 \todo[inline]{We argue that a new performance metric called \textit{goodput}, emphasizing transmission efficiency over volume, should replace the traditional concept of \textit{throughput}.}

\begin{itemize}
\item We introduce the Search-in-Memory (SiM) chip, a standalone flash memory chip minimally adopted from existing chips to realize on-chip equality tests.  SiM features a versatile SIMD interface with two primitives: search and gather command (\cref{sec:main-method}). This interface makes SiM adaptable for various data-bound operations, offering flexibility and applicability to different scenarios.

\item Maintaining data integrity is a significant challenge for NAND-flash-based on-chip computing. To address this, we propose the  ``Optimistic Error Correction'', which optimizes the common case of no errors in Single-Level Cell (SLC) pages, while providing a fallback solution for rare corner cases (\cref{sec:implementation}).

\item We introduce several system integrations, from general data structures like B+Tree, which is used in many systems, to supporting database analytical queries, to demonstrate SiM's generalizability and flexibility. (\cref{sec:integrations}).

\end{itemize}

\section{Background and Motivations}

\subsection{SSD's Parallelism}

\begin{figure}[t]
    \centering
    \includegraphics[width=\linewidth]{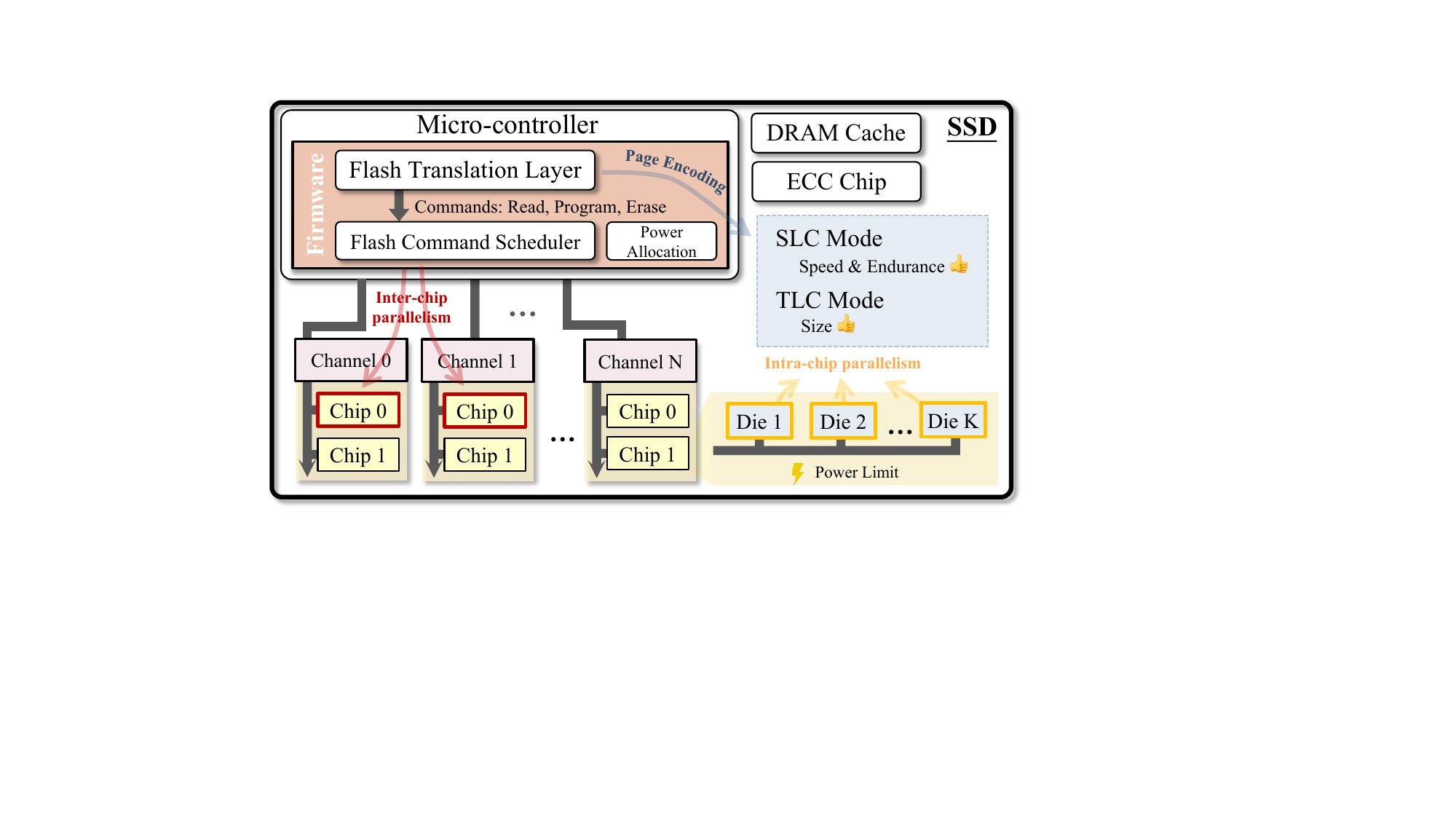}
    \caption{SSD Architecture}
    \label{fig:ssd-overview}
    \vspace{-1.6em}
\end{figure}

As shown in \cref{fig:ssd-overview}, an Solid-State drive (SSD) is made up of multiple flash memory chips that communicate with a central controller via high-speed data channels.  A chip has several dies, each can simultaneously conduct memory operations. Modern SSDs' impressive I/O bandwidth is the result of parallel operations across multiple chips (i.e., \textit{inter-chip parallelism}) and the activation of multiple components within a single chip (i.e., \textit{intra-chip parallelism}). However, the degree of parallelism has a physical limit.  Heat dissipation is becoming increasingly difficult, even in data centers, as the density of modern flash memory chips increases.  Too many parallel operations can result in electric currents that exceed the hardware power budget.

\begin{figure}[t]
    \centering
    \includegraphics[width=.8\linewidth]{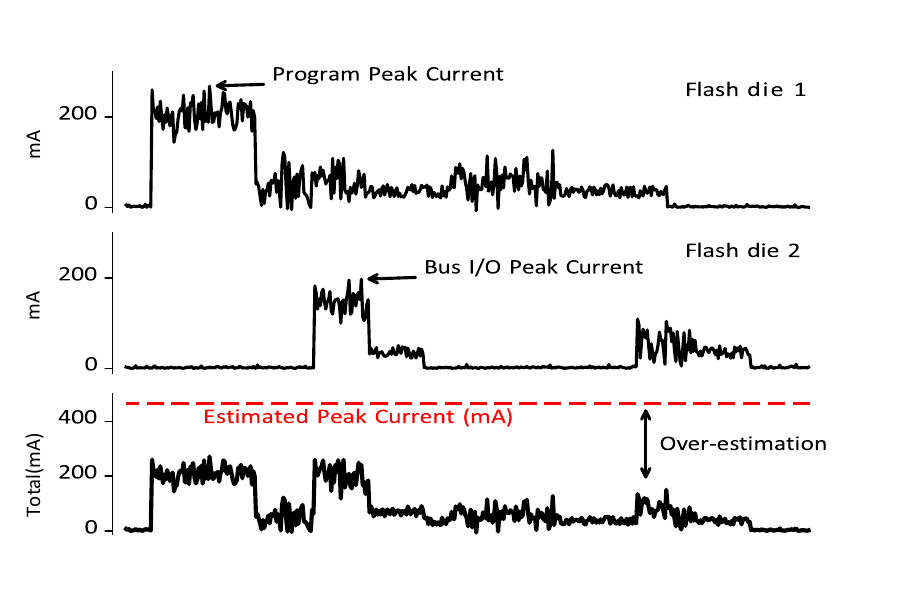}

    \vspace{-1em}
    \caption{Conceptual illustration of current consumption in a NAND Flash chip}
    \label{fig:peak-current}
    \vspace{-1.3em}
\end{figure}

\subsection{Bus I/O can limit SSD's parallelism}
As shown in \cref{fig:peak-current}, a flash memory command consumes varying amount of current throughout various phases (I/O transfer phase, the read/program phase, and the status phase).  To simplify power management, many controllers represent the peak current consumption of a command as its overall current usage \cite{ppmcache20,ppm22,ppm23}.  This ensures that the total current consumption of the entire chip does not exceed the power budget when multiple commands are executed concurrently.  On the other hand, if the aggregate peak currents is anticipated to exceed the power budget, the controller must restrain from dispatching further commands even if the target flash die is idle.  Lowering the peak current of a flash command is therefore critical for ensuring efficient power allocation and parallelism. 

As SSDs' capacity increases, more data must be moved in and out, increasing the demand for higher I/O bandwidth \cite{skhynix23}. The increased bandwidth requirement is often fulfilled by increasing the I/O clock rate, but such an approach can easily make the I/O phase to become the phase in a flash command that draws the peak current.  
For instance, transferring a 16 KiB page at a clock frequency of 1.6GHz can consume up to 50\% of a chip's maximum power budget \cite{ppmcache20}. 

Performance scaling through continually increasing the I/O clock rate is not sustainable and there is a need to fundamentally reduce the bandwidth demand.  In fact, as we will show in this paper, a decrease in I/O bandwidth does not always result in lower application performance. By filtering out unnecessary data transfer at its source, it is possible to operate I/O buses at a reduced clock rate while preserving the application-perceivable throughput.  This paper aims to enable such a filtering at low cost.

\subsection{Capacity and Metadata Scaling Must Go Hand-in-hand}  \label{sec:slc-mlc}
Recently, improvements in 3D-NAND Flash memory technology have made it possible to stack more than 300 layers of memory cells \cite{skhynix23}, each cell storing multiple bits.  This increases SSD's capacity to unprecedented levels.    However, without proportionate scaling of metadata storage, the efficiency of retrieving the increased volume of data will be seriously compromised.

SSDs use Single-Level Cell (SLC) and Triple-Level Cell (TLC) modes to encode data and metadata differently in order to meet the specific needs of data and metadata storage.  The speed and durability of SLC mode--—which stores one bit per cell—--make it the preferred method for storing metadata.  TLC mode---which stores three bits per cell---is used for data storage because it has a higher storage density.  

\begin{figure}[t]
    \centering
    \begin{minipage}[t]{\linewidth}
      \centering
      \includegraphics[width=.95\linewidth]{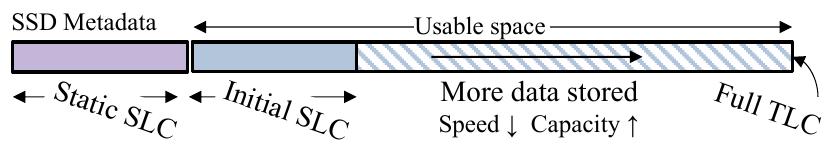}
      \vspace{-.07in}
      \caption{Commercial SSD}
      \label{fig:overall-ssd-a}
    \end{minipage}\hfill
    \vspace{.1in}
    \begin{minipage}[t]{\linewidth}
      \centering
      \includegraphics[width=.95\linewidth]{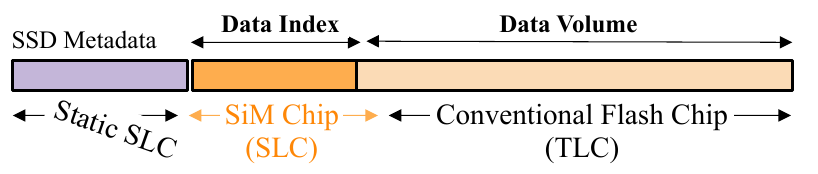}
      \vspace{-.07in}
      \caption{SiM-enhanced SSD}
      \label{fig:overall-ssd-b}
    \end{minipage}
  
    \vspace{-1em}
  \end{figure}

\cref{fig:overall-ssd-a} depicts the architecture of a typical commercial SSD. A small section of the memory cell encoded in SLC is used to store internal metadata or a write buffer, while the remaining memory cell encoded in TLC is used to store user data.  The user data section can transition between SLC and TLC depends on capacity usage.  Data are stored in SLC if user utilizes less than advertised capacity.  As more capacity is used, the SSD controller transparently converts the SLC-encoded data into TLC.  However, such an implicit hybrid model does not guarantee that the user's metadata will be accessed optimally.

In this paper, we propose allocating a portion of the user-visible capacity to store data indexes, as depicted in \cref{fig:overall-ssd-b}.  We implement the index storage with the Search-in-Memory chip in SLC mode.  Although our model has a lower total capacity than using TLC mode for the entire user visible capacity, it provides better metadata access performance and endurance.

\subsection{The Case for a New Chip Optimized for Data Indexing} \label{sec:the-case-for-new-chip}
There must be a compelling case for designing a new hardware solution because it might bring a huge enineering cost.
Data indexing is frequently the first step in querying large data systems such as file systems, databases, and search engines for narrowing down the search space. 
The process of executing a key query on a typical database index is as follows.
First, an in-memory index structure is queried to locate the leaf index pages. These leaf index pages can be, for example, the leaf node of a B-Tree or a bucket in a hash table. Then, the leaf index page is searched to locate the corresponding entry.
These  indexes are so large that they must be stored on SSDs and loaded into host memory on demand before the CPU can search the query key in the array of candidate entries in the index pages.  The search is usually performed using either SIMD or binary search.  However, transferring a large number of index pages between SSD and host memory for matching is usually the performance bottleneck.  It also consume significant amount of I/O bandwidth and power. 

The I/O bottleneck in data indexing between the SSD and host has led to the development of various near-storage processing solutions, which conduct data matching in the SSD controller's CPU \cite{kevin20,kvssd19}. However, we argue that instead of loading the vast amount of candidateentries  into general-purpose processors to match with a small query  key; we should reverse the I/O direction by shipping the query key to where the candidate entries are stored.  Several Processing-in-Memory proposals have used this approach \cite{pim12,tseng20,ice22}.  However, many proposals incorporate a processing element (PE) into the memory array or a specialized pattern matching accelerator \cite{pif22} in the peripheral circuit, increasing design complexity and manufacturing costs.

This paper demonstrates the feasibility of adapting the existing design of NAND flash chips to enable on-chip index search. We find that index searches can utilize the existing logic gates within a flash memory chip's peripheral circuits, reducing the need for substantial additional hardware investments. This approach repurposes hardware initially intended for core data storage functionalities. For instance, the registers and logic gates within each page buffer, originally designed for the encoding and decoding of multiple bits within a memory cell, can be repurposed to execute bit-serial matches. Similarly, the page-wide counter, initially devised for verifying data programming, can be adapted for the aggregation of match results.  This strategic repurposing of existing circuits introduces new indexing capabilities while maintaining the original functionalities and without significantly affecting the chip's area or power budget.

\section{Search-in-Memory (SiM)} \label{sec:main-method}
We introduce the Search-in-Memory (SiM) chip, which integrates vectorized data matching into NAND flash memory. This allows data-bound operations to be executed directly within the SSD, eliminating the need to transfer index pages to the CPU. Rather than viewing index pages as opaque data, SiM treats the page content as an array of fixed-width data.

SiM offers a generic SIMD interface, featuring two primary commands: \textit{search} and \textit{gather}. The \searchcmd{} compares an input key with the data array in the index page, generating a matching bitmap.  Subsequently, the \gathercmd{} uses this bitmap to extract specific data chunks within an index page, bypassing non-matching data. This targeted approach reduces the bandwidth waste and excessive energy often linked with full page-sized I/O transfers.

\todo[inline]{
Furthermore, the \searchcmd{} can be batched, returning multiple bitmaps, thereby amortizing the search cost.  The bitmaps can then be merged for a subsequent batched \gathercmd{}.
}

\begin{figure}[t]
  \centering
  \includegraphics[width=\linewidth]{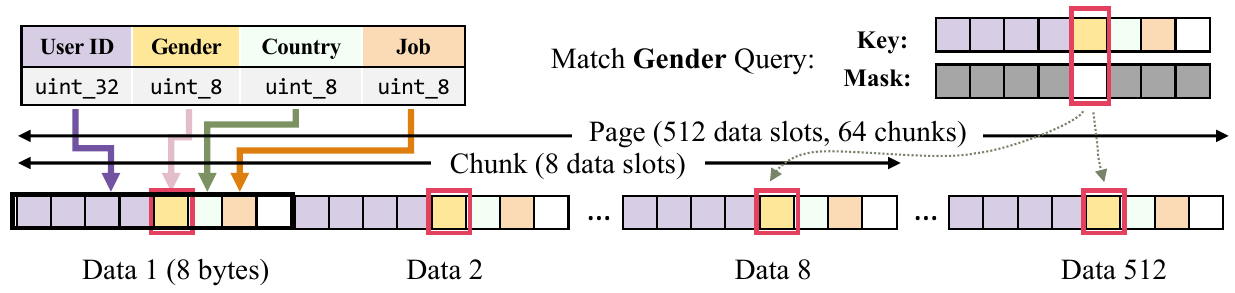}
  \caption{Page format and data encoding}
  \label{fig:page-format}
  \vspace{-1.em}
\end{figure}

\subsection{SiM Page Format}
As shown in \cref{fig:page-format}, SiM recognizes a data page as an array of 8-byte data slots, a format central to many index structures, like the leaf node of a B+Tree or an external hash table's bucket.  Thus, a 4KiB page corresponds to an array with 512 data slots\footnote{Throughout the rest of the paper, we use 4KiB as the logical page size}. When a \searchcmd{} is performed, the chip matches the 8-byte query key with these slots, returning a 512-bit bitmap as the match result.

To reduce wiring overhead in the \gathercmd{} implementation, we group every eight data slots into a \textit{chunk}. This chunk serves as the minimal data transfer unit. Optionally, users can treat the first chunk as the page header, using it to store metadata, a practice common in many B+Tree implementations.

\subsection{SiM Command Format}

SiM's \searchcmd{} consists of the target page address and two 64-bit arguments: a query key and a mask. The mask facilitates the comparison of specific bit ranges, ignoring other positions as ``don't care''. In SiM-indexed relational database tables, where each row corresponds to an 8-byte key and data columns are encoded at specific bit ranges, the mask aids in isolating a specific column for matching. \cref{fig:page-format} demonstrates this by encoding rows into 8-byte data and querying based on the gender value, while masking unrelated columns. This command format flexibility enables SiM to support diverse queries through the \textit{BitWeaving} technique \cite{bitweaving13}, which is widely used in database systems to enable high parallelism.

\todo[inline]{

The search command has a variant akin to the conventional \textit{Read Cache} command to enable pipelined search, called \cachecmd{}.  When the controller needs to search another flash page within the same block, it can directly send a \cachecmd{} to the chip, specifying the target page address.  The target page can be loaded from the memory without affecting the data accessed by concurrent \texttt{search} commands because they are stored in distinct registers.
Once all necessary search commands have been executed, the controller can send a \texttt{page-close} command to release the registers used by the current page, allowing bit matching to be performed on the page loaded by the \cachecmd{}.

}

SiM's \gathercmd{} resembles the \texttt{gather} SIMD instruction for the CPU: it uses a 64-bit index bitmap to indicate the desired chunks within an index page to read (a page contains 64 chunks).  Compared to transmitting the entire page, the \gathercmd{} can significantly reduce the volume of I/O transmission.   

\subsection{Storage and Match Mode}
SiM ensures compatibility with existing flash memory chips and preserves their high-density storage functionality by introducing minimal additional hardware. It operates in two modes: \textit{Match Mode} and \textit{Storage Mode}. A flash memory page can function in both modes, but their interpretations differ.

In \textit{Storage Mode}, the flash memory chip is solely responsible for storing data. It does not interpret the page content. This mode emphasizes high storage density and I/O bandwidth. Consequently, it typically stores multiple bits per memory cell, and the I/O bus operates at a high clock rate.

In contrast, \textit{Match Mode} prioritizes efficient data retrieval. It stores only one bit per cell (Single-Level Cell, or SLC) to ensure data reliability, and the I/O bus operates at a lower clock rate. This mode does not compromise latency because on-chip matching significantly reduces the amount of data transfer required.

SiM dynamically switches between the two modes based on operational requirements. It utilizes \textit{Match Mode} for foreground indexing operations, taking advantage of its efficient data retrieval capabilities. On the other hand, it employs \textit{Storage Mode} when writing new data and performing background maintenance, leveraging its high storage density and I/O bandwidth.

\section{Implementation}\label{sec:implementation}

\begin{figure}[t]
    \centering
\includegraphics[width=.9\linewidth]{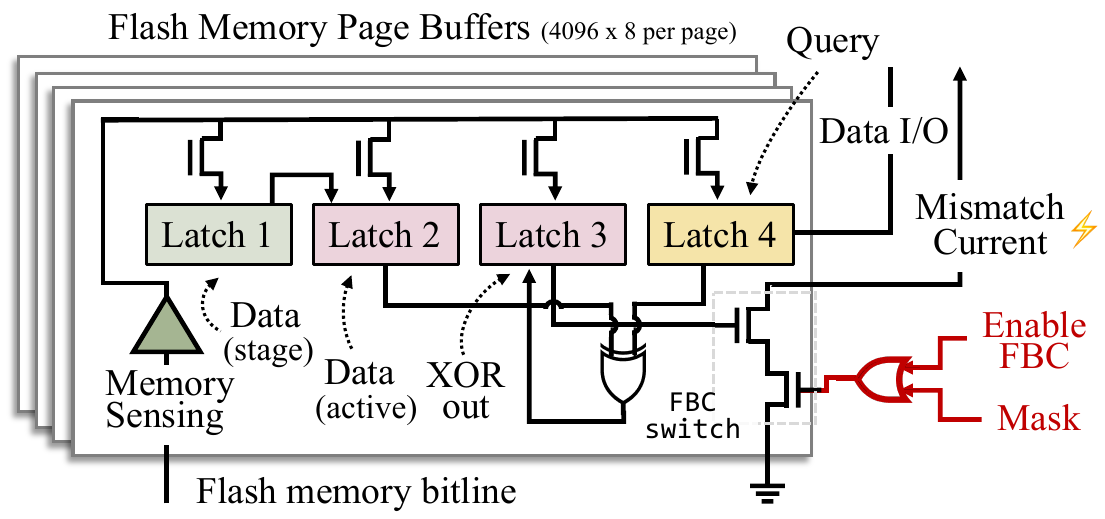}
    \caption{Page Buffer (extension to existing structure marked in red)}
    \label{fig:page-buffer}
    \vspace{-1em}
\end{figure}

\subsection{Extending Existing Circuit} \label{page_buffer_extension_implementation}

Each NAND flash memory plane contains a set of page buffers, each associated with a memory bitline, for reading a page  from the memory array. \cref{fig:page-buffer} illustrates the typical structure of a NAND flash memory page buffer (PB), equipped with multiple data latches\footnote{A data latch stores a single bit. Encoding and decoding 3-bit storage require three latches and an XOR gate.} and an XOR gate.

\begin{figure}[t]
    \centering
\includegraphics[width=.92\linewidth]{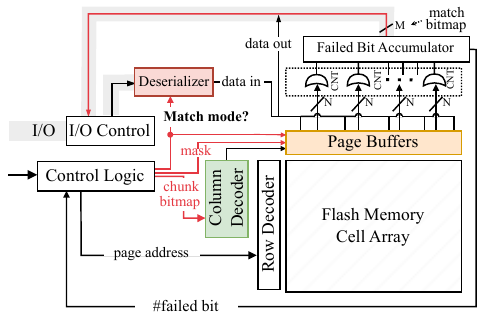}
    \caption{SiM's Chip-level Design}
\label{fig:sim-circuit}
\vspace{-.15in}
\end{figure}

SiM utilizes the \texttt{XOR} gate for bit matching, in conjunction with the Failed Bit Counting (FBC) circuitry\footnote{SSDs store data by injecting electric charges into flash memory cells until they reach a predetermined charge level. After every program operation, the cell states are verified
and recorded in Latch 3.  A one-bit denotes a mismatch, releasing a small current. The FBC sums these currents, determining if the misprogrammed cells exceed a set limit.  Every 64 PBs are grouped and all currents from the group's PBs are combined using an analog counter, with the current magnitude indicating the count value \cite{fbc18, fbc20}.}. Query key is loaded into Latch 4 and XORed with the memory content stored in Latch 2. The XOR result is stored in Latch 3, where a one-bit signifies a mismatch. SiM's core data unit, including its query key size and mask size, is 8 bytes (or 64 bits) to align with the FBC's PB group structure, where every 64 bitlines form a match group.  A non-zero count in a PB group indicates a mismatch.
Moreover, we add an \texttt{OR} gate to each PB. This allows reading from Latch 2 either when FBC is activated during data programming in \textit{Storage Mode} or when the current query's mask bitmap has an active bit in the respective bit position in \textit{Match Mode}.
\cref{fig:sim-circuit} shows SiM's chip design, incorporating a new signal, \texttt{match mode}, to switch between \textit{Storage Mode} and \textit{Match Mode}.

\subsection{On-chip Matching Workflow} \label{on-chip-matching}

\todo[inline]{\textbf{Remove cache mode}: On-chip matching is initialized as follows.  Firstly, the \opencmd{} command is executed to load data from the flash memory into Latch 1. If the previous bit-matching round has not received the \texttt{page-close} command, indicating the presence of active data in Latch 2, we enter cache mode. In cache mode, the new data is held in Latch 1 without impacting the contents of other latches. Once the \texttt{page-close} command is issued to release the active page, the newly read page transitions to normal mode for the subsequent round of bit matching.

If no active matching is in progress, SiM enters normal mode. First, it performs the \textit{Optimistic Error Correction} to ensure data integrity (refer to \cref{sec:indexing-optimistic-ecc}). Then, the data is prepared as an operand for the subsequent matching operation by transferring it to Latch 2. Once the initialization is complete, SiM is ready for bit matching.
}

The controller initiates on-chip matching using the \opencmd{}, which specifies the target page address. Upon receiving this, SiM loads the target page from flash memory into Latch 1.  Since SiM permits simultaneous memory reading and bit matching from a previous round, active data from the previous round might still be in Latch 2.  If so, the newly-read data is held in Latch 1 until the \texttt{page-close} command moves it to Latch 2.  The next round then begins with \textit{Optimistic Error Correction} for data integrity (refer to \cref{sec:indexing-optimistic-ecc}).

After the initialization, SiM can receive multiple \searchcmd{} for batch matching. Each \searchcmd{} activates the deserializer to duplicate and forward the 64-bit query key to Latch 4 of each page buffer. In the next clock cycle, Latch 2 and Latch 4 contents are XORed, and the result is stored in Latch 3. A replicated 64-way mask signal, representing the 64-bit mask of the query, is linked to every page buffer. If both the XOR result in Latch 3 and the mask signal are active, a small current flows through the FBC switch, signaling a \textit{mismatch}.
The FBC's analog counter then aggregates the 64 match signals. If there's a mismatch, it emits a non-zero value, which is identified using a 1-bit voltage comparator. A bitmap of size $M = 512$, denoting match results from $M$ PB groups (with each group having 64 PBs), is generated. These results are stored in latches for synchronization and later transferred to the I/O bus\footnote{Unlike normal data transfer, which sends approximately equal numbers of zero and one bits, the bitmap from the SiM chip mostly comprises zero bits due to the typically low number of matches. This sparsity reduces power consumption during data transmission over modern I/O bus protocols operating in \textit{Low-Tapped Termination}, like NV-LPDDR4 \cite{ltt21}, which consumes power only when transmitting one bits.}.

SiM performs a \gathercmd{} as follows. First, the target page is loaded from the flash memory into the L1 latch. Next, the column decoder deserializes the 64-bit index bitmap, converting it into the entire page, and then sequentially transmits the selected chunks onto the I/O bus. It is common for a \searchcmd{} to be immediately followed by a \gathercmd{}. In such cases, since the page content is already loaded into the page buffers, the \gathercmd{} can initiate data transmission without delay.

\subsection{Data integrity}
\subsubsection{Data randomization}

In modern SSDs, it is a common practice to randomize the stored data to ensure data reliability. This randomization process involves XORing the data bits with a deterministically generated random bit stream, which is derived from a seed determined by the page address. When reading a page, the data is de-randomized using the same procedure to recover the original data values. In SiM, the query key is randomized within the \textit{deserializer} using the same seed that was used to randomize the target page. Since the random stream is cancelled out when XORed twice, we can perform bit matching in the page buffer without de-randomizing the target data page. Unlike conventional randomization, we initialize the seed for each chunk using the chunk address. This enables us to de-randomize non-contiguous chunks in the \gathercmd.

\subsubsection{Optimistic Error Correction} \label{sec:indexing-optimistic-ecc}
In order to perform on-chip matching without transmitting the full page to the SSD controller, we adopt an optimistic approach of sampling a few bytes at the beginning of the page for errors. This approach is based on two rationales. Firstly, a recent work in in-flash computing \cite{flashcosmos22} has characterized real chips and found that the SLC pages we adopt, which store one bit per cell, exhibit no errors for extended periods of time. Secondly, another recent work has demonstrated the feasibility of sampling a portion of a page to determine its overall stability \cite{sentinel20}.

Our optimistic approach is as follows. Before writing a logical page to the flash memory, we prepend a verification header to verify data integrity during subsequent page reads. This verification header includes the current timestamp and an 8-byte predetermined magic number. Additionally, we prepend an 8-byte CRC checksum calculated over the first chunk and the two aforementioned fields.

When the \opencmd{} loads the page content from the flash memory, both the verification header and the first chunk are transmitted to the controller. The controller verifies the chunk using the CRC checksum. If a mismatch is detected, the controller initiates a full page read to retrieve the entire page from the page buffer. The page is then processed by a dedicated ECC chip, similar to a normal page read. If an uncorrectable error is detected, the controller adjusts the sensing voltage using the magic number and performs read-retries up to a specified maximum number of times  \cite{sentinel20}.

Our optimistic error correction approach optimizes the common case of error absence in SLC pages while providing a fallback solution for corner cases. Additionally, if the age of the page, indicated by the write timestamp in the verification header, exceeds a safety margin, the page is also read out for error correction and placed in a refresh queue to be rewritten at a later time, ensuring data reliability.

\subsubsection{Concatenated Error Correction}

In addition to the verification header, we also assign a 4-byte ECC parity to each chunk, which is checked in the controller upon loading. The chunk-level ECC is stored alongside the page-level ECC parity. This arrangement forms a \textit{concatenated code}, a classic technique for enhancing data reliability \cite{forney1965concatenated}. In our case, this arrangement enables the \gathercmd{} to perform fine-grained error correction without the need to load the entire page to the controller.

\subsection{Hardware Overhead}
We add the mask signal to each PB to control the FBC switch in match mode and an \texttt{OR} gate to enable the FBC in data programming in storage mode. We also modify the column decoder to transmit specific chunks within a page and adjust the deserializer to distribute the input data across all page bits. Given that modern NAND flash chips support reading specific portions of a page (i.e., \textit{Random Data Out}) and generate test data patterns for reliability tests \cite{bist22}\yc{\cite{bist17}}, our modifications to the column decoder and deserializer are minimal. Considering the page buffer and decoder account for under 9\% of the total chip area \cite{pif22}, we estimate that SiM adds around 3\% to the overall area overhead. 

\subsection{Batch Matching} \label{sec:batch-matching}

SiM offers the capability of batch matching to maximize the utility of a page read from the flash memory (the page read latency accounts for the largest portion in the overall on-chip matching process, so conducting multiple matching can amortize the page read latency).  
We implement a deadline-based command scheduler to evaluate the effectiveness of this approach. Each command is associated with a deadline upon submission. The scheduler holds the submitted commands in a queue until their respective deadlines expire. At that point, the scheduler searches for other commands in the queue that target the same page and submits them together as a batch.  We evaluate the scheduler in \cref{batch-cim-eval}.

\section{System Integrations} \label{sec:integrations}
This section demonstrates how SiM's versatile interface makes it possible to integrate it into various data-intensive systems.

\subsection{Database Primary Index} \label{integration:db-primary-index}
\begin{figure}[t]
    \centering
    \includegraphics[width=\linewidth]{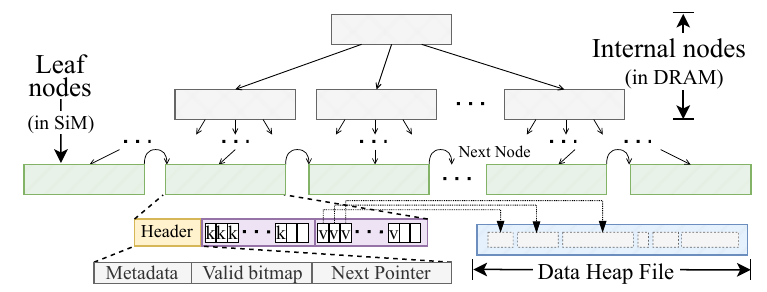}
    \caption{SiM-enhanced Database Primary Index}
    \label{fig:btree-integration}
    \vspace{-.18in}
\end{figure}

The \textit{Primary index} in a relational database maps the primary key of a table to a pointer indicating the storage location of the corresponding data row.  It is usually implemented with a B+Tree, as shown in \cref{fig:btree-integration}.  The internal nodes of the B+Tree can usually fit within the DRAM, while the leaf nodes often require on-demand reading from disk \cite{btree11}. A leaf node page typically begins with a header that stores metadata, including a validity bitmap, counters for empty slots, compression information, and sibling pointers. Following the header is a compact array of keys and values.

The arrangement of keys and values within the leaf nodes is essential to efficient search of query key on the CPU. For instance, keys can be stored contiguously or sorted to facilitate SIMD parallel search and binary search, respectively. The leaf nodes can be directly stored in SiM, effectively replacing the process of on-demand disk I/O and in-CPU search with a search command to SiM. A leaf node can span multiple SiM pages. For example, an 8KiB leaf node can store its key array in one SiM page and its value array in another SiM page. A leaf node search involves a search command that targets the first page, followed by a gather command that targets the second page. These two commands can be internally pipelined to reduce latency. Storing keys and values separately, as opposed to storing them in the same page, increases the parallelism of key search and prevents unnecessary loading of the value when a key is not found.

\begin{table}[ht]
    \centering
    \caption{SiM- versus non-SiM-based Primary Index}
    \label{tab:indexing-comparison-sim-with-without}
		\resizebox{0.5\linewidth}{!}{\footnotesize
    \begin{tabular}{@{}lll@{}}
      \toprule
      & SiM    & Without SiM           \\ \midrule
      Total I/O & 128 B  & 8192 B                \\
      Bus Freq  & 40 MHz & 1600 MHz \cite{onfi41}              \\
      Current   & 11 mA \cite{ymtc-gen2-datasheet}  & 152 mA \cite{ppmcache20}       (13x)    \\
      Energy     & 63 nJ  & 1400 nJ     (22x)     \\
      Latency   & 3.2 $\mu$s & 5.1 $\mu$s         (1.6x) \\ \bottomrule
    \end{tabular}

    }
  
\vspace{-.07in}
  \end{table}

\cref{tab:indexing-comparison-sim-with-without} presents a back-of-the-envelopment comparison of the worst-case energy consumption and latency in data transfer between a conventional disk-based B-Tree and a SiM-based B-Tree. The comparison focuses solely on the data transfer from the flash memory chip's page buffer to the SSD controller, excluding transfer to the host OS. 
In the absence of SiM, the entire key and value pages must be read, resulting in an I/O size of 8 KiB. However, with SiM, the \searchcmd{} sends a 64-byte bitmap, while the \gathercmd{} sends a 64-byte chunk. The flash chip operates in \textit{Match Mode} with a bus clock frequency of 40 MHz, whereas without SiM, the clock frequency defaults to 1600 MHz for higher bandwidth. Consequently, the peak current of the high-speed bus is thirteen times greater than that of the low-speed bus. Energy consumption is 22 times higher without SiM. However, the latencies of the two approaches are comparable. 
This comparison demonstrates that SiM's data reduction improves energy efficiency and performance due to enhanced \textit{goodput}.

\begin{figure}[t]
    \centering
    \includegraphics[width=.9\linewidth]{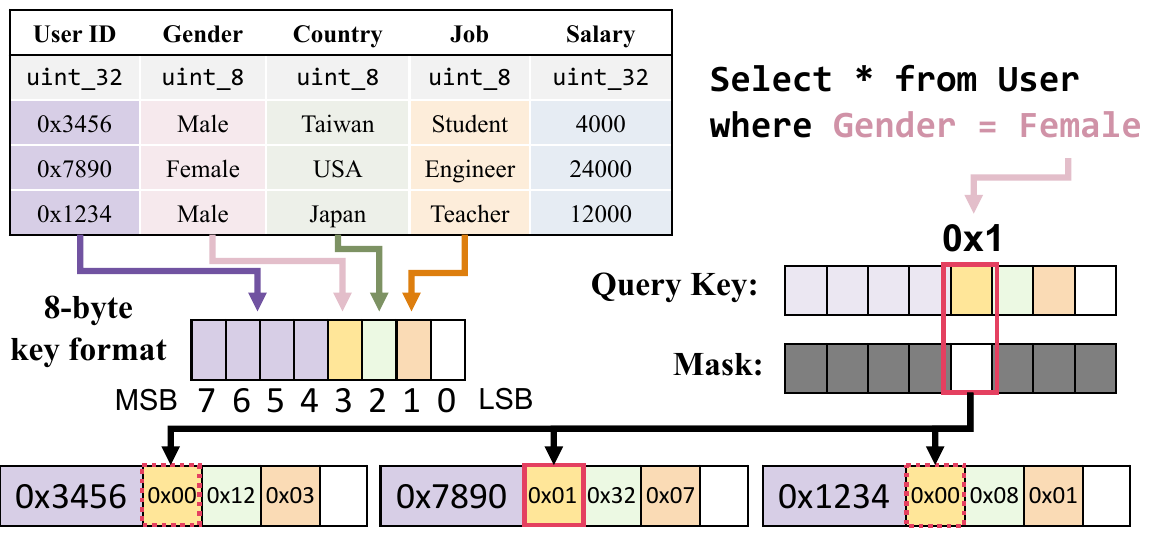}
    \caption{Example Encoding of a Table for SiM}
    \label{fig:secondary-index-integration}

    \vspace{-.2in}
\end{figure}

\subsection{Database Secondary Index}
A secondary index is a data structure in a database that maps the values of one or multiple columns to the primary key of a table, enabling efficient retrieval of rows based on specific column values without the need for a full table scan. Figure~\ref{fig:secondary-index-integration} illustrates an example user table with its secondary index stored on SiM. Following the key encoding scheme used in MySQL \cite{myrocks20}, each row is transformed into an 8-byte key, and the encoded keys are stored compactly in a SiM page. To perform a query that retrieves all female users, the target value (e.g., female represented by \texttt{0x01}) is encoded into the query key, and a mask is constructed based on the position of the target column. SiM produces a match bitmap, allowing us to retrieve the user IDs of the matched keys using a \gathercmd over the same page.

\subsection{Database Range Queries}
A range query in a database table aims to find all $k$ such that $U > k \geq L$.  SiM narrows the search space in two steps.  First, it decomposes the range query into upper-bound and lower-bound queries. The upper-bound query $U > k$ is transformed into $2^{\lceil \log_2(U)\rceil}-1 > k$, where $2^{\lceil \log_2(U)\rceil}$ corresponds to the smallest value larger than $U$ that is a power of two.  The lower-bound query $k \geq L$ is processed by transforming it into an upper-bound query of ``$k < L$'' and then applying a bitwise \texttt{NOT} operation to the obtained bitmap. The final result of the range query is obtained by performing a bitwise \texttt{AND} operation between the two sub-queries.

\begin{figure}[th]
    \centering
    \includegraphics[width=.9\linewidth]{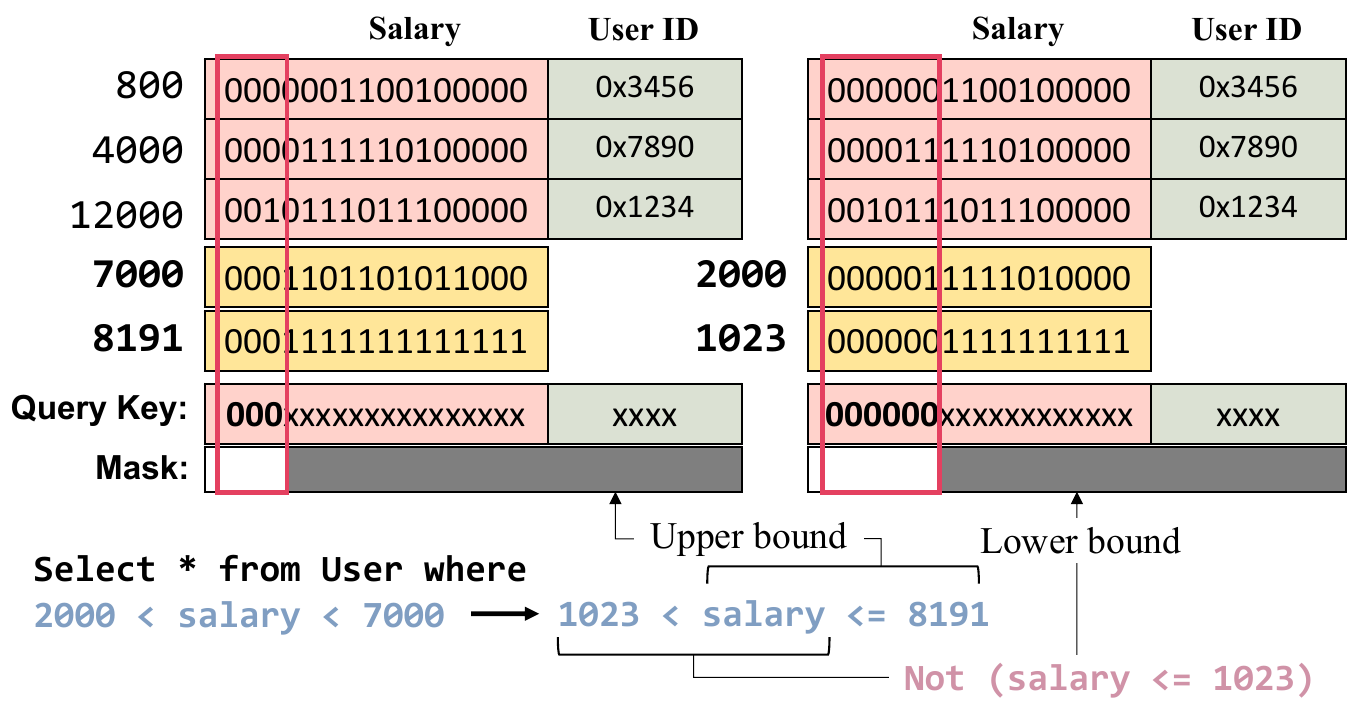}
    \caption{Example Usage of SiM: Filtering a Secondary Index}
    \label{fig:db-filter-integration}
\vspace{-.06in}
\end{figure}

\cref{fig:db-filter-integration} illustrates a secondary index for finding users with a specific range of the salary column in the same table. The salary and user ID are encoded in big endian into an 8-byte key, resembling the bit-sliced index used in analytical databases \cite{bitslice15}. The range query ``\texttt{select * from User where 2000 < salary < 7000}'' is decomposed into upper-bound and lower-bound queries.
In \cref{fig:db-filter-integration}, the upper-bound query is transformed into ``\texttt{salary <= 8191}''. By examining the most significant bits, we determine that the \nth{0} to \nth{2} bits of both 7000 and 8191 are zero. Any integer with the \nth{0} to \nth{2} bits being zero is guaranteed to be smaller than 8191. The \nth{0} to \nth{2} bits in the query key are set to zero while masking out the rest of the bit positions. The search returns the bitmap \texttt{110} because 800 and 4000 satisfy the conditions.  The lower-bound query is transformed into ``\texttt{Not (salary <= 1023)}'', which returns the bitmap \texttt{011}.  Combining the two sub-queries give the final result: \texttt{010}.

Although the result encompasses more elements than the actual range, it effectively reduces the search space for subsequent fine-grained filtering.
We make the design choice of only implementing exact equality matching over exact range search because this allows us to repurpose existing circuitry in the hardware implementation without the need for additional circuits, as further discussed in \cref{on-chip-matching}.  There is no bit dependencies in exact equality matching.  Thus, the matching can be finished in one pass, and the inter-bit wiring cost can be saved.

Nevertheless, users have the flexibility to conduct multi-pass comparisons to achieve their desired level of confidence.  This can be achieved by masking out the previously-compared MSB bit region and recursively compare the masked-out number.
Furthermore, in the field of data analytics, precise results are often unnecessary.  When the keys are uniformly distributed, our approximate range query can have low error rates \cite{bitslice15}.

\subsection{Redistributing Data}
Many data structures used in disk-based systems partition and redistribute data to improve performance. For instance, when a B-Tree or extendible hash table becomes full, it splits a full node or bucket into two. LSM-Trees perform compaction when a level reaches its capacity. Log-structured data structures require periodic garbage collection to free up space. In database systems, the join operation combines data from multiple tables using a hash table, which necessitates partitioning the dataset to ensure efficient data access during queries. These operations involve reading data from disk and rearranging them in memory. Data redistribution can result in high temporary memory usage and CPU spikes, especially in log-structured storage where data for the same partition can be scattered across multiple files. This can cause significant performance issues for frontend user services.  Because of its significance, there have been calls for specialized hardware acceleration \cite{xengine20}.   However, with SiM, data redistribution can be performed incrementally by keyspace partitioning.  Partitioning the key space using a specific bit slice from the key, similar to a radix tree, allows us to locate a particular partition using the \searchcmd{} and collect the data using the \gathercmd{}. By gathering one partition at a time, we can avoid loading data that do not belong to the specific partition, effectively reducing the I/O and memory overhead.

\section{Experiments}\label{sec:evaluation}

  \begin{table}[]
    \centering
    \caption{Hardware Parameters}
    \label{tab:hardware-conf}
    \begin{tabular}{ll}
    \hline
    \multicolumn{2}{l}{3D NAND flash chip parameters}                        \\ \hline
    \multicolumn{1}{l|}{(Channel, Package)}           & (8, 1)               \\
    \multicolumn{1}{l|}{(Die, Plane, Block, Page)}    & (2, 1, 32, 128)      \\
    \multicolumn{1}{l|}{(Read, Program, Erase)}       & (16 µs, 80 µs, 1 ms) \\
    \multicolumn{1}{l|}{(Flash Page Size, Cell type)} & (4 KiB, SLC)         \\
    \multicolumn{1}{l|}{SiM Clock Cycle}              & 10                   \\
    \multicolumn{1}{l|}{SiM Clock Frequency}          & 33 MHz               \\ \hline
    External I/O                                      &                      \\ \hline
    \multicolumn{1}{l|}{Interface}                    & PCIe Gen 3               \\
    \multicolumn{1}{l|}{Bus Width}                    & 128                  \\
    \multicolumn{1}{l|}{Bus Clock}                    & 250MHz               \\ \hline
    Internal I/O Bus                                  &                      \\ \hline
    \multicolumn{1}{l|}{Interface}                    & NV-DDR3 (ONFi 4.x)   \\
    \multicolumn{1}{l|}{Bandwidth (Match mode)}       & 80 MT/s              \\
    \multicolumn{1}{l|}{Bandwidth (Storage mode)}     & 800 MT/s             \\
    \multicolumn{1}{l|}{Bus Width}                    & 8                    \\ \hline
    \multicolumn{2}{l}{Power Settings}                                       \\ \hline
    \multicolumn{1}{l|}{(Bus Voltage, NAND Voltage)}  & (1.2V, 3.3V)         \\
    \multicolumn{1}{l|}{Bus Active / Idle Current}    & 5mA /  10uA          \\
    \multicolumn{1}{l|}{NAND Read/Program Current}    & 25mA, 25mA           \\
    \multicolumn{1}{l|}{SiM Current}                  & 2.5mA               
    \end{tabular}
  \vspace{-.14in}
    \end{table}

\subsection{Experimental setup}
\subsubsection{Hardware}
We implement the \searchcmd{} and \gathercmd{} by defining two new NVMe commands.  We encode the SiM-specific payloads in NVMe's vendor-specific Dataset Management (DSM) opcode and  extend NVMe's kernel driver to parse the new command formats.
We prototyped and simulated SiM on Amber \cite{simplessd:18}, a high-fidelity SSD emulator. \cref{tab:hardware-conf} shows the hardware parameters we used.  The I/O bandwidth for SiM's match command is configured 80 MT/s (NV-DDR3's timing mode 1), which is 10\% of the typical bandwidth for full-page I/O.  This setting can lower the I/O bus's operation current\footnote{While higher timing modes typically incur higher operational current as in \cref{tab:indexing-comparison-sim-with-without}, we set the I/O bus current consumption of the baseline also to 5mA assess the inclusion of advanced power optimization \cite{ltt21}.} and the peak power.  Thanks to the reduced I/O volume, the latency and the energy is minimally effected.  I/O requests are scheduled in a First-Come-First-Serve manner, but a deadline-based scheduler is also evaluated in \cref{batch-cim-eval}.

\subsubsection{Data structure}
\begin{figure}[h]
  \centering
  \includegraphics[width=.6\linewidth]{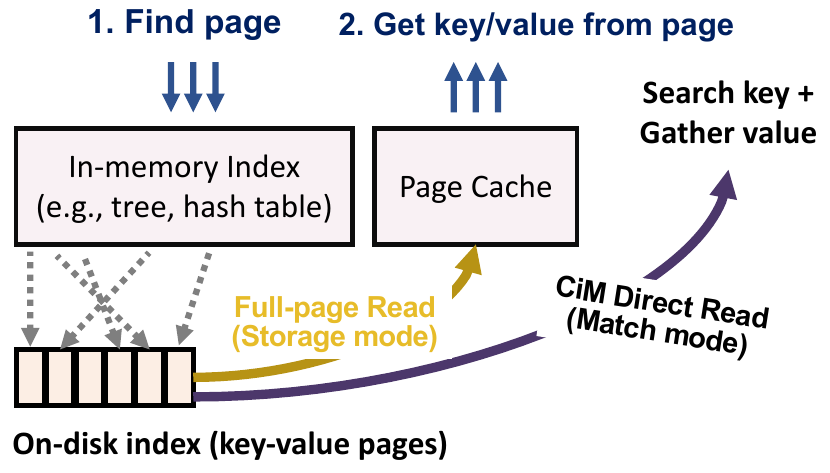}
  \caption{SiM and baseline setup in the experiment}
  \label{fig:exp-arch}
  \vspace{-.1in}
\end{figure}

We create a generic index that consists of an in-memory top-level index and a collection of disk pages, each containing a compact array of key-value pairs.  The top-level index maps a key to its on-disk page, as shown in \cref{fig:exp-arch}.  If it is implemented as a B-Tree, the on-disk pages correspond to the leaf nodes.  If implemented as a hash table, on-disk pages correspond to hash buckets.  The on-disk page is then loaded into the operating system's page cache, from which the value can be searched. 
The on-disk index is set to 650 MiB, taking 65\% of the simulated SSD's capacity.  We ensure that there is enough spare space to prevent SSD space reclamation, allowing for a more focused evaluation. 

\subsubsection{Baseline Setup}
Our baseline is the traditional CPU-centric architecture, which reads entire pages from disk and stores frequently accessed pages in the page cache. SiM, on the other hand, bypasses the page cache by sending a \searchcmd{} to the on-disk page address of a key specified by the in-memory index, determining the key's position in the page, and retrieving the desired values using a \gathercmd{}.  As we will see later, bypassing the page cache effectively frees it up for other uses, such as write buffering.   Thanks to the small transmission size, SiM communicates with the host OS entirely through NVMe's command interface (i.e., MMIO) and bypass the conventional DMA procedures. 
Note that this paper lacks direct comparison with ParaBit \cite{parabit21} and CoX-PM \cite{pif22} due to differing application scenarios and difficulties in accurately reproducing their proprietary environments.

\subsubsection{Workloads}
We customize the Yahoo! Cloud Serving Benchmark (YCSB) \cite{ycsb10,zonelife23} and subject the index to various query distributions and read/write patterns to evaluate it across various application scenarios. Using Linux's CGroup, we downscale the page cache size to various ratios of the on-disk index size. The term \emph{Cache Coverage} refers to this ratio. For example, a Cache Coverage of 50\% indicates that the page cache size is 325 MiB, which is 50\% of the on-disk index size (650 MiB).  A Cache Coverage of 0\% indicates that caching is disabled.

We begin collecting statistics only after the initial data has been loaded into the SSD and the workload has run for 30\% of its designated length to ensure the system reaches steady state.
We disable periodic cache flushing of dirty pages to better understand the systems' sensitivity to varying cache sizes. 

\begin{table}[h]
  \caption{Query Concentration in Different Distributions}
  \label{tab:query_distribution}
  \centering
  \begin{tabular}{@{}lcccc@{}}
  \toprule
              & 1st & 2nd & 3rd & 4th \\ \midrule
  Uniform & 0.03\% & 0.03\% & 0.03\% & 0.03\% \\ 
  Skewed ($\alpha = 0.5$) & 0.23\% & 0.12\% & 0.11\% & 0.07\% \\ 
  Very Skewed ($\alpha = 0.9$) & 17.00\% & 2.54\% & 1.53\% & 1.08\% \\ \bottomrule
  \end{tabular}
\end{table}

\subsubsection{Query Distribution}

\cref{tab:query_distribution} illustrates query concentration across different distributions. In many online services, it's common for a small number of queries to dominate the workload. This phenomenon is modeled using Zipf's distribution for both skewed ($\alpha = 0.5$) and very skewed ($\alpha = 0.9$) scenarios, alongside a uniform distribution for comparison.  The uniform distribution shows an even spread, whereas the very skewed distribution ($\alpha = 0.9$) shows a significant dominance of the top queries, with the most frequent accounting for 17\% of the total workload.

\section{Results}

\begin{figure*}[h]
  \centering
  \begin{minipage}[t]{0.43\linewidth}
    \centering
    \includegraphics[width=\linewidth]{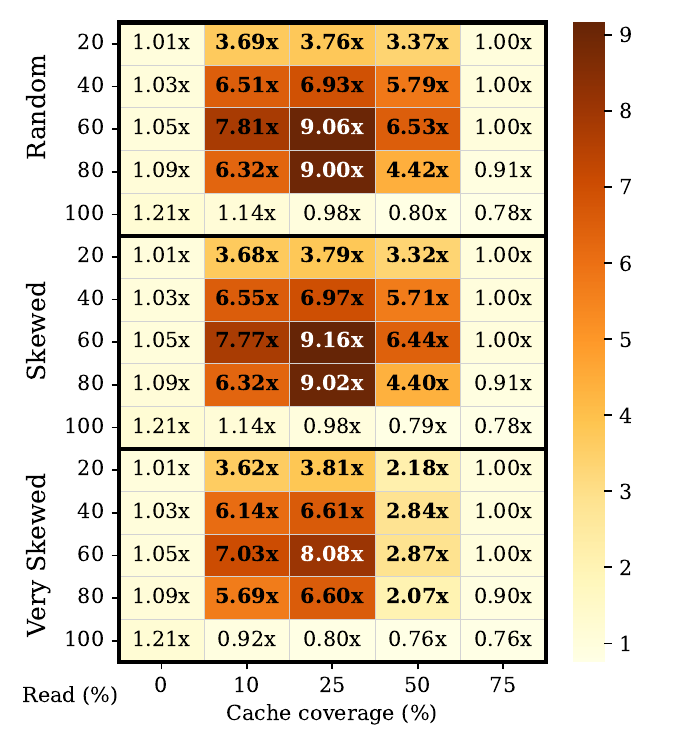}
    \vspace{-.25in}
    \caption{SiM's query-per-second speedup over baseline}
    \label{fig:qps-compare}
  \end{minipage}\hfill \begin{minipage}[t]{0.43\linewidth}
    \centering
    \includegraphics[width=\linewidth]{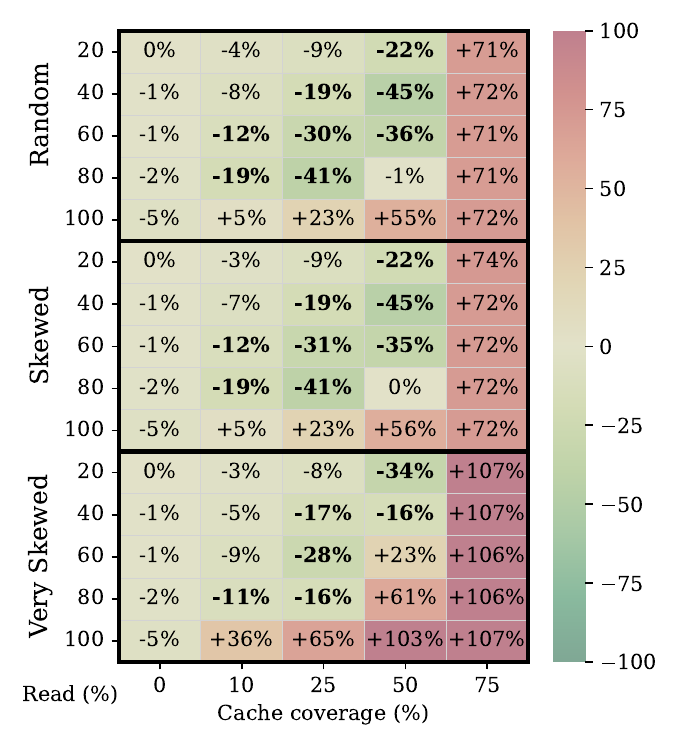}
    \vspace{-.25in}
    \caption{SiM's energy consumption over baseline}
    \label{fig:energy-compare}
  \end{minipage}
    \vspace{-.2in}
\end{figure*}

\subsection{Overall Speedup} \label{result:overall-speedup}
\cref{fig:qps-compare} displays SiM's overall speedup in terms of query per second (QPS)\footnote{A workload's QPS is a measure of its throughput calculated as the number of queries divided by the completion time (excluding the first 30\% of queries designated as warmup period).} compared to the baseline.  The Y axis represents the varying percentage of read requests in the workload.  100\% indicates a completely read-only workload, while 20\% indicates a write-intensive workload.  The X axis depicts different cache coverage.  0\% disables the page cache and directs all I/O to the SSD.  10\% and 25\% would be the typical configuration for real-world system to balance performance and hardware cost. 
We make several observations from \cref{fig:qps-compare}: 

\begin{itemize}
  \item The baseline supported with cache performs 8-20\% better than SiM in read-only workloads.  This is to be expected, given that SiM bypasses the page cache and requires additional cycles for on-chip matching, whereas the baseline may avoid I/O by searching the pages stored in the page cache directly.  One possible solution is to send search commands to the same page in batches to amortize the latency of reading from NAND flash memory---a technique evaluated in \cref{batch-cim-eval}.  Another option is to cache the retrieved keys.  This fine-grained management can make better use of cache space than traditional page-level caching, but it can significantly complicate index designs.  
  \item SiM outperforms 3X to 9X in write-intensive workloads.  This is because SiM does not use read caching, so the cache can be used for write buffering.  Because writes are significantly more expensive than reads on SSDs, increasing write buffering can improve overall performance and extend SSD lifespan.  This is consistent with the design of many modern database engines, such as RocksDB, where read requests bypass page cache to avoid prematurely evicting dirty pages from cache.
  \item When cache coverage is zero, all I/O goes directly to the SSD, and the locality difference in query distribution has no effect on performance.  When cache coverage is high (75\%), SiM has few performance advantages over the baseline because the cache is large enough to absorb page updates that would otherwise be evicted under low cache coverage.
\end{itemize}

\subsection{Energy consumption}
\cref{fig:energy-compare} compares SiM's energy consumption with the baseline.  This analysis favors the baseline because it only considers the energy consumption of the NAND flash chip, ignoring the energy consumption of the CPU and DRAM, which are difficult to accurately characterize.  It also equalizes the baseline's bus I/O current consumption with SiM's to incorporate recent power optimization for high-frequency I/O bus \cite{ltt21}.  
Even with these assumptions, SiM still reduces energy consumption by 10\% $\sim$ 45\% at typical cache coverage levels (10\% $\sim$ 50\%).  A cache coverage of 75\% is only a reference as it does not account for the significant DRAM energy consumption required to provide a large memory space.  SiM's ability to lower write traffic is what accounts for the lower energy use, as further explored in \cref{fig:program-count-compare}.

SiM's ability to reduce read I/O also contributes to energy savings. In contrast to the baseline, which sends complete key and value pages (each 4 KiB) to the host OS via the PCIe bus, SiM only transmits the result bitmap (64B) from the key page and the necessary chunk (64B) from the value page for a random point query, where only one chunk is needed. This strategy decreases data transmission over the PCIe bus by 64 times.  In the internal I/O bus, SiM needs to transfer another 256B for integrity verification upon page open, but this still reduces I/O by 21 times.  This is why, despite using a 10-times slower bus timing mode, SiM can reduce I/O transmission delay and bus active time by 2.1 times.  

SiM's I/O reduction lowers queuing delays and shortens the SSD's active period.  These benefits effectively offset the additional energy consumed by SiM for on-chip matching operations.

\begin{figure*}[h]
  \centering
  \begin{minipage}[t]{0.43\linewidth}
    \centering
    \includegraphics[width=\linewidth]{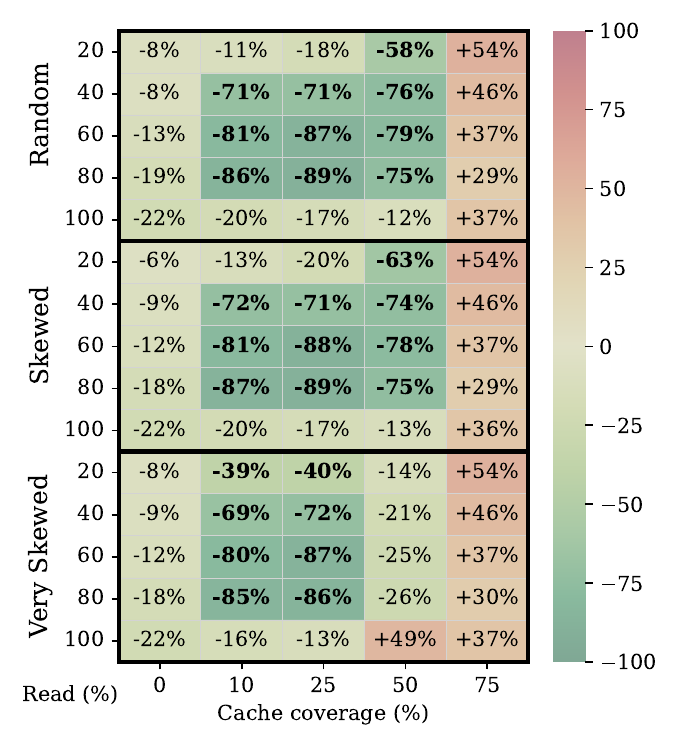}
    \vspace{-.25in}
    \caption{SiM's Median Read Latency Reduction}
    \label{fig:lat-median-compare}
  \end{minipage}\hfill \begin{minipage}[t]{0.43\linewidth}
    \centering
    \includegraphics[width=\linewidth]{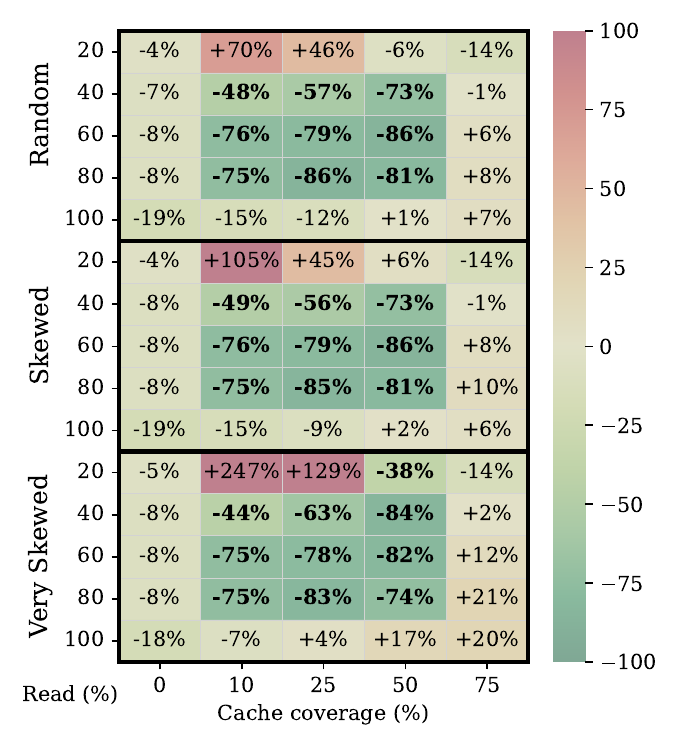}
    \vspace{-.25in}
    \caption{SiM's Tail Read Latency Reduction}
    \label{fig:lat-99th-compare}
  \end{minipage}

  \vspace{-.2in}
\end{figure*}

\subsection{Read Latency}

\cref{fig:lat-median-compare} compares SiM's median read latency reduction to the baseline. This reduction varies from 30\% to 89\% across workloads, whether skewed or uniformly distributed.  Note that this analysis inherently favors the CPU-centric baseline, as we omit the CPU's search time for locating the target key after key pages have been loaded into host OS's memory. In contrast, for SiM, we include the latency incurred by on-chip matching operations. Despite this discrepancy that could advantage the baseline, SiM still demonstrates superior latency improvements. 

\begin{figure}[h]
  \centering
  \begin{subfigure}[t]{0.9\linewidth}
    \centering
    \includegraphics[width=.8\linewidth]{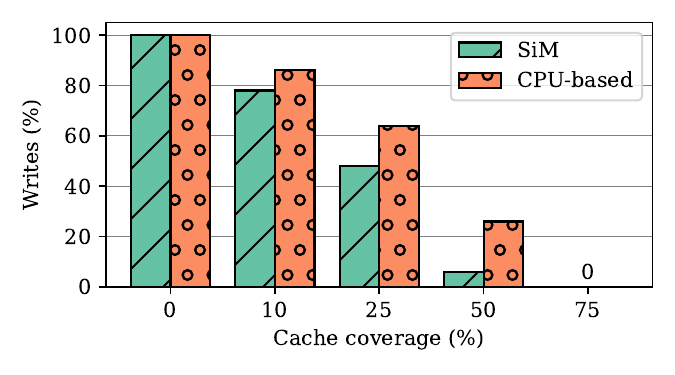}
    \vspace{-.6em}
    \caption{Amount of Writes Relative to No Caching}
    \label{fig:program-count-compare}
  \end{subfigure}\vspace{.6em}
  \hfill \begin{subfigure}[t]{0.9\linewidth}
    \centering
    \includegraphics[width=.8\linewidth]{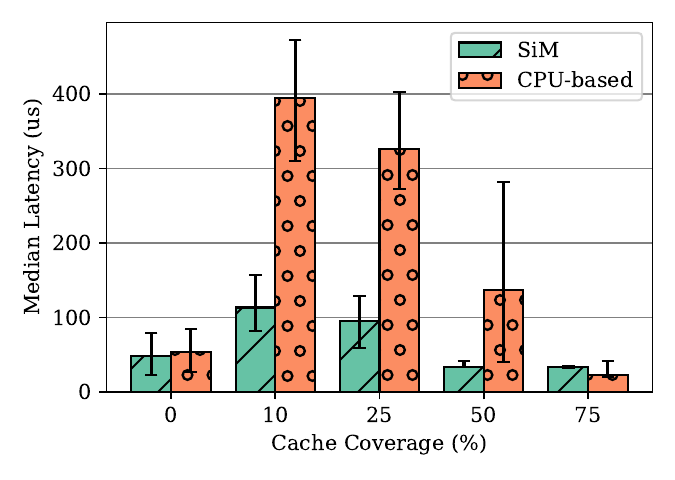}
    \vspace{-.6em}
    \caption{Median Read Latency}
    \label{fig:lat-median-compare-zoom}
  \end{subfigure}\caption{ Detailed comparison at 40\% Read, Random Dist.}
  \label{fig:lat-and-program-comparison}
  \vspace{-.2in}
\end{figure}

In read-only workloads, SiM outperforms the baseline particularly when the baseline is allocated less cache.  This can be attributed to the longer I/O transmission of the full page transfer.
\cref{fig:lat-median-compare-zoom} zooms in on the comparison of median read latencies between SiM and the baseline under a random query distribution and a 40-60 read-write workload. Here, error bars denote the \nth{25} and \nth{75} percentiles, with SiM demonstrating narrower error bars. This suggests a more consistent response time, which is crucial for services directly interacting with users.

In write-intensive workloads, SiM has lower read latency than the baseline in mid-range cache coverage where the write set size exceeds the cache capacity.  In this case, new writes can evict both clean and dirty pages.  Clean page eviction degrades read performance due to cache misses, whereas dirty page eviction causes lengthy queueing delays for read operations.  SiM's cache bypass strategy, as discussed in \cref{result:overall-speedup}, alleviates this effect.

\begin{figure*}[h]
  \centering
  \begin{minipage}[t]{0.4\linewidth}
    \centering
    \includegraphics[width=.97\linewidth]{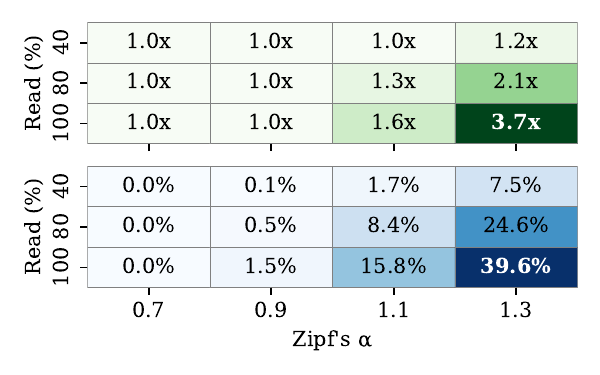}
    \vspace{-.11in}
    \caption{QPS speedup of Batch Submission (top) and Merge probability (bottom)}
    \label{fig:qps-vs-merge}
  \end{minipage}\hfill \begin{minipage}[t]{0.6\linewidth}
    \centering
    \includegraphics[width=.97\linewidth]{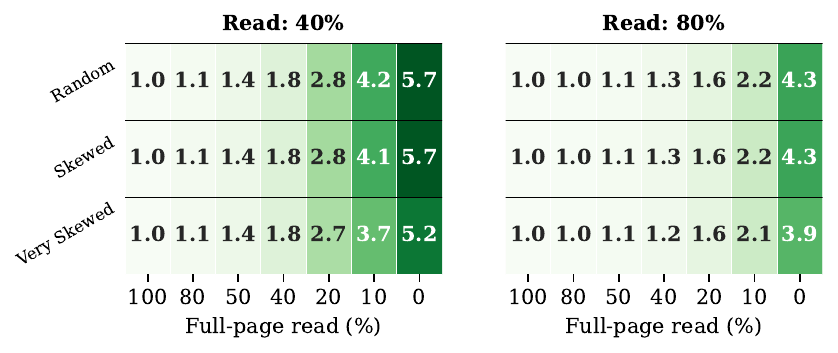}
    \vspace{-.11in}
    \caption{QPS Speedup versus Full-page Read Ratio}
    \label{fig:full-page-sensitivity}
  \end{minipage}

  \vspace{-.2in}
\end{figure*}

\subsection{Tail Read Latency}  \label{sec:tail-latency}

\cref{fig:lat-99th-compare} presents the tail read latency (\nth{99} percentile) improvements SiM achieves over the baseline. Although the variability between the \nth{25} and \nth{75} percentile latencies is less for SiM, in rare cases, SiM may exhibit significantly higher latency compared to the baseline, particularly in workloads where read requests are infrequent and highly skewed. Closer examination reveals differing write patterns: the baseline experiences consistent write activity, whereas SiM may face sporadic peaks in write demand. This is attributed to SiM's page cache being primarily composed of dirty pages from data writes. Consequently, initiating a new write could trigger a chain reaction of writing back dirty pages, potentially delaying read requests substantially. In contrast, the baseline system's page cache contains some clean pages fetched from the SSD, which can be evicted immediately to buffer data writes, avoiding such corner cases.

To mitigate this issue, implementing an I/O scheduler that gives priority to reads over writes could prevent read starvation. Alternatively, preempting writes in favor of reads—a strategy proposed for ultra-low-latency SSDs \cite{ull:18}—could also be effective. Future research should explore replacing our current First-Come-First-Serve I/O scheduling with more sophisticated strategies to assess their impact on reducing tail latency.

\subsection{Batch CiM submission} \label{batch-cim-eval}
\cref{sec:batch-matching} introduces a deadline scheduler aiming to reduce NAND flash memory's read latency by batching \searchcmd{} for identical pages. Each \searchcmd{} is assigned a deadline of 4$\mu$s, which constitutes 25\% of the 16-microsecond read latency for SLC memory. The upper heatmap of \cref{fig:qps-vs-merge} presents the query-per-second improvement when using the deadline scheduler, compared to SiM's performance without it. The lower heatmap indicates the probability that a query will target the same page as another unexpired query in the scheduler. 
As the concentration of queries increases, indicated by a rising Zipf's $\alpha$ value, the probability of multiple queries targeting the same page increases, resulting in a 3.7-fold boost in throughput at $\alpha=1.3$ for purely read-only workloads.  However,  such an $\alpha$ value is way beyond what a normal workload would exhibit.  Setting a longer expiration time can also improve throughput, but at the expense of prolonged latency.   We conclude that the deadline scheduler is ineffective for low-latency SSDs because the overhead outweighs the benefits.

\subsection{Sensitive Analysis on Full-page Read Ratio}

While SiM excels in precise data retrieval, the need for full-page reads remains crucial. For instance, indices in read-heavy analytic databases require summing data across entire pages. Similarly, the write-optimized LSM-Tree index, while needing efficient support for random point queries, also necessitates compaction—a garbage collection process that reads indices in full pages for merging. This leads us to assess how variations in the volume of full-page reads affect overall performance across different query distributions and in both read- and write-dominant workloads. \cref{fig:full-page-sensitivity} illustrates the relative query-per-second speedup of SiM compared to the baseline where all reads are full-page (on the left-most side of the X axis). Observe that as the proportion of SiM reads within the workload increases, so does performance. This effect is evident in both read- and write-dominant scenarios, though more markedly in the latter.  On the other hand, the influence of varying query distributions on this trend appears minimal.

\section{Related Works} \label{sec:related-works}

Numerous research efforts have been made on minimizing data transfers through early data filtering, which can be broadly classified into near-storage computing approaches—--such as SmartSSD or custom circuits attached to flash memory controllers \cite{instorage:23}--—and on-chip computing approaches like SiM. Near-storage computing reduces I/O between the host and SSD, whereas on-chip computing reduces data movement from within the SSD itself.  On-chip approaches can be analog-based or digital-based.  Analog approaches, such as Tseng et al. \cite{tseng20}, are well-suited for error-tolerant applications like machine learning but fall short for the precise data matching required in indexing. Digital approaches, like Parabit \cite{parabit21} and Flash Cosmos \cite{cosmos22}, use the existing flash memory sensing mechanisms for bulk bitwise operations such as AND and OR across flash pages.  SiM also utilizes the existing page buffer circuits but has a different programming model. Unlike Parabit and Flash Cosmos, where both operands are page-sized and  must be pre-programmed into the same NAND block prior to the computation, SiM operates with a small query and a page for comparison, making it more efficient to deal with small, dynamically-loaded operands.

CoX-PM \cite{pif22}  incorporates error correction and pattern matching circuits into the NAND flash memory. SiM, on the other hand, chooses not to perform on-chip error correction due to its complexity, instead relying on Optimistic Error Correction on SLC pages and the SSD controller's existing ECC chips. SiM also opts not to evaluate complex pattern matching in hardware, instead using software to decompose complex queries into elementary instructions that are cheaper to implement in hardware.  ICE \cite{ice22} integrates 8-bit integer multiplication into the peripheral circuits for on-chip vector matching. Unlike CoX-PM and ICE, SiM strives to repurpose existing circuits and minimize additional circuits to reduce hardware testing costs and accelerate adoption.

\section{Conclusion}
 This paper introduced the Search-in-Memory (SiM) chip, a novel solution aimed at overcoming the bottleneck in data indexing through on-chip data matching. SiM introduces simple yet versatile commands for fine-grained data searching and gathering. These commands, despite their simplicity, enable complex, data-intensive operations found in various data structures to be accelerated. SiM's command structure allows for cost-effective implementation with minimal modifications to existing circuit designs.  Furthermore,  SiM can be combined with readily available high-capacity NAND flash memory chips to create a hybrid SSD that effectively realize the principle of data-metadata separation.

SiM has undergone extensive testing under a variety of workload and system constraints.  Evaluation shows up to 9X speedup in write-intensive workloads and up to 45\% energy savings due to reduced read and write I/O and better utilization of host's cache space.  SiM reduces median and tail read latency by up to 89\% and 85\%, respectively.  As a future work, we aim to integrate SiM technology into actual key-value and relational database systems to enhance their efficiency in garbage collection and range queries.  Developing a hardware prototype is also planned.

\bibliographystyle{IEEEtran}

\bibliography{phd3-remove-doi}

\end{document}